\documentclass[twocolumn]{article}

\usepackage{arxiv}

\usepackage[utf8]{inputenc} 
\usepackage[T1]{fontenc}    
\usepackage{hyperref}       
\usepackage{url}            
\usepackage{booktabs}       
\usepackage{amsfonts}       
\usepackage{nicefrac}       
\usepackage{microtype}      
\usepackage{lipsum}
\usepackage{graphicx}
 \usepackage{amsmath}
\usepackage{amssymb}
\usepackage{dcolumn}

\usepackage{tipa}
\graphicspath{{imgs/}}
\title{Muiltiscale modeling of electrical conductivity of R-BAPB polyimide + carbon nanotubes nanocomposites}

\author{S. V. Larin \\
Institute of Macromolecular Compounds, Russian Academy of Sciences,\\
V.O. Bol'shoi pr. 31, 199004 St. Petersburg, Russian Federation\\
\And
S. V. Lyulin \\
Institute of Macromolecular Compounds, Russian Academy of Sciences,\\
V.O. Bol'shoi pr. 31, 199004 St. Petersburg, Russian Federation\\
\And
P. A. Likhomanova\\
National Research Center "Kurchatov Institute", 123182, Moscow, Russia \\
\texttt{likhomanovapa@gmail.com} \\
\And
K. Yu. Khromov\\
National Research Center "Kurchatov Institute", 123182, Moscow, Russia and\\
Moscow Institute of Physics and Technology (State University), 117303, Moscow, Russia\\
\texttt{khromov\_ky@nrcki.ru}\\
\And
A. A. Knizhnik\\
Kintech Laboratory Ltd., 123182, Moscow, Russia and\\
National Research Center "Kurchatov Institute", 123182, Moscow, Russia\\
\And
B. V. Potapkin\\
Kintech Laboratory Ltd., 123182, Moscow, Russia and\\
National Research Center "Kurchatov Institute", 123182, Moscow, Russia}

\begin{document}
	
	\twocolumn[{
	
\maketitle
\begin{abstract}
The electrical conductivity of the polyimide R-BAPB polymer filled with single-wall carbon nanotubes (CNT) with chirality (5,5) is modeled using a multi-scale approach. The modeling starts with molecular dynamics simulations of time-dependent  fluctuating atomic configurations of polymer filled CNTs junctions. Then the atomic positions obtained in the first step are used to perform fully first-principles microscopic calculations of the CNTs junctions contact resistances using the  Green's function based quantum transport technique. And finally, those contact resistances are supplied as an input to a 
statistical calculation of a CNTs ensemble conductivity using a Monte Carlo percolation model.
The results of the first-principles calculations show a very strong dependence of the polymer filled CNTs junctions contact resistance on the geometry of CNTs junctions, including an angle $\varphi$ between nanotubes axes and the positions of polymer atoms around CNTs. Incorporating into the percolation model  this strong dependence as well as  CNTs agglomeration,  pushed the calculated values of electrical conductivity just above the percolation threshold below 0.01 S/m, which is within the experimental range for composites with various base polymers. Possible mechanisms for further reduction of composites conductivity are discussed.
\end{abstract}

	}]

\section*{Introduction}
Polymer materials, while possessing some unique and attractive qualities, such as low weight, high strength, resistance to chemicals, ease of processing, are for the most part
insulators. If methods could be devised to turn common insulating polymers into conductors, that would open great prospects for using such materials in many more areas than they are
currently used. These areas may include organic solar cells, printing electronic circuits, light-emitting diodes, actuators, supercapacitors, chemical sensors, and biosensors \cite{Long_2011}.

Since the reliable methods for carbon nanotubes (CNT) fabrication had been developed in the 1990s, growing attention has been paid to
the possibility of dispersing CNTs in polymers, where CNTs junctions may form a percolation network and turn insulating polymer into a good conductor when a percolation threshold is overcome.
An additional benefit of using such polymer/CNTs nanocomposites instead of intrinsically conducting polymers, such as polyaniline \cite{Polyaniline}
for example, is that dispersed CNTs, besides providing electrical conductivity, enhance
polymer mechanical properties as well.

CNTs enhanced polymer nanocomposites have been intensively investigated experimentally, including composites conductivity \cite{Eletskii}. As for the theoretical research in this area, the results are more modest. 
If one is concerned with nanocomposite conductivity, its value 
depends on many factors, among which are the polymer type, CNTs density, nanocomposite preparation technique, CNTs and their junctions geometry, a possible presence of defects
in CNTs and others. Taking all these factors into
account and obtaining quantitatively correct results in modeling is a very challenging task since the resulting conductivity is formed at different length scales: at the microscopic level it is influenced by the
CNTs junctions contact resistance and at the mesoscopic level it is determined by  percolation through a network of CNTs junctions. Thus a consistent multi-scale method for the modeling of conductivity,
starting from atomistic first-principles calculations of electron transport through CNTs junctions is necessary.

Due to the complexity of this multi-scale task, the majority of investigations in the area are carried out in some simplified forms, this is especially true for the underlying part of the modeling: determination
of CNTs junction contact resistance. For the contact resistance  either experimental values as in \cite{Soto_2015} or the results of phenomenological Simmons model as in  \cite{Xu_2013, 
Yu_2010, Jang_2015, Pal_2016} are usually taken, or even an arbitrary value of contact resistance reasonable by an order of magnitude may be set \cite{Wescott_2007}. In \cite{Bao_2011, Grabowski_2017} the tunneling probability through a CNT junction is modeled using a rectangular potential barrier and the quasi-classical approximation.

The authors of \cite{Castellino_2016} employed an oversimplified two-parameter   expression for contact resistance, with these parameters fitted to the experimental data.
The best microscopic attempt, that we are aware of, is using the semi-phenomenological tight-binding approximation for the calculations of contact resistance \cite{Penazzi_2013}. But in \cite{Penazzi_2013} just
the microscopic part of the nanocomposites conductivity problem is addressed, and the conductivity of nanocomposite is not calculated. Moreover, in \cite{Penazzi_2013} the coaxial CNTs configuration is only considered,
which is hardly realistic for real polymers.

Thus, the majority of investigations are concentrated on the mesoscopic part of the task:
refining a percolation model or phenomenologically taking into account different geometry peculiarities of CNTs junctions. Moreover, comparison with experiments is missing in some publications on this topic.
Thus, a truly multi-scale research, capable of providing quantitative results comparable with experiments, combining fully first-principles calculations of contact resistance on the microscopic level with a percolation model on 
the mesoscopic level seems to be missing. 

In our previous research \cite{comp_no_pol}, we proposed an efficient and precise method for fully first-principles calculations of CNTs contact resistance and combined it with a Monte-Carlo statistical percolation model to
calculate the conductivity of a simplified example network of CNTs junctions without polymer filling. 
In the current paper, we are applying the developed approach  to the modeling of conductivity of the CNTs enhanced polymer polyimide R-BAPB. 

\begin{figure}[b]
\includegraphics[width=8cm]{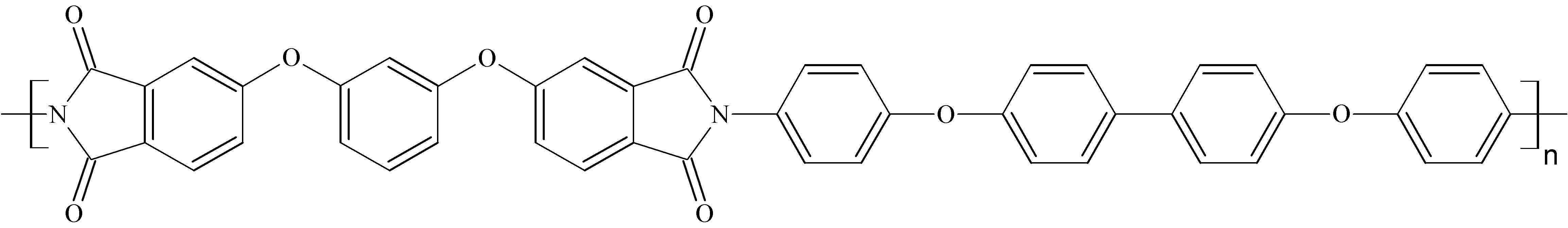}
\caption{The chemical structure of R-BAPB polyimide.} 
\label{fig_RBAPB-struct}
\end{figure}

R-BAPB (Fig. \ref{fig_RBAPB-struct}) is a novel polyetherimide synthesized using 1,3-bis-(3$'$,4-dicarboxyphenoxy)-benzene (dianhidride R) and 4,4$'$-bis-(4$''$-aminophenoxy)diphenyl (diamine BAPB). It is thermostable polymer with extremely high thermomechanical properties (glass transition temperature $T_g= 453-463$~K, melting temperature $T_m= 588$~K, Young's modulus $E= 3.2$~GPa) \cite{Yudin_JAPS}. This polyetherimide could be used as a binder to produce composite and nanocomposite materials demanded in shipbuilding, aerospace, and other fields of industry. The two main advantages of the R-BAPB among other thermostable polymers are thermoplasticity and crystallinity. R-BAPB-based composites could be produced and processed using convenient melt technologies.

Crystallinity of R-BAPB in composites leads to improved mechanical properties of the materials, including bulk composites and nanocomposite fibers. It is well known that carbon nanofillers could act as nucleating agents for R-BAPB, increasing the degree of crystallinity of the polymer matrix in composites. As it was shown in experimental and theoretical studies
\cite{Yudin_MRM05, Larin_RSCADV14, Falkovich_RSCADV14,Yudin_CST07}, the degree of crystallinity of 
carbon nanofiller enhanced R-BAPB may be comparable to that of bulk polymers.

Ordering of polymer chains relative to nanotube axes could certainly influence a conductance of the polymer filled nanoparticle junctions. However, it is expected that such influence will depend on many parameters, including the structure of a junction, position, and orientation of chain fragments on the nanotube surface close to a junction, and others. Taking into account all of these parameters is a rather complex task that requires high computational resources for atomistic modeling and ab-initio calculations, as well as complex analysis procedures. Thus, on the current stage of the study, we consider only systems where the polymer matrix was in an amorphous state, i.e. no sufficient polymer chains ordering relative to nanotubes were observed.

\section*{Description of the multiscale procedure} 

The modeling of polymer nanocomposite electrical conductivity is based on a multi-scale approach, in which different simulation models are used at different scales. For the electron transport in polymer composites with a conducting filler, the lowest scale corresponds to the contact resistance between tubes. The contact resistance is determined at the atomistic scale by tunneling of electrons between the filler particles via a polymer matrix, and hence, analysis of contact resistance requires knowledge of the atomistic structure of  a contact. Therefore, at the first step, we develop an atomistic model of the contact between carbon nanotubes in a polyimide matrix using the molecular dynamics (MD) method. This method gives us the structure of the  intercalated polymer molecules between carbon nanotubes for different intersection angles between the nanotubes. One should mention, that since a polymer matrix is soft, the contact structure varies with time and, therefore, we use molecular dynamics to sample these structures. 

Based on the determined atomistic structures of the contacts between nanotubes in the polymer matrix we calculate electron transport through the junction using electronic structure calculations and the formalism of the Green’s matrix. Since this analysis requires first-principles methods, one has to reduce the size of the atomistic structure of a contact to acceptable values for the first-principles methods, and we developed a special procedure for cutting the contact structure from MD results. First-principles calculations of contact resistance should be performed for all snapshots of an atomistic contact structure of MD simulations, and an average value and a standard deviation should be extracted. In this way, one can get the dependence of a contact resistance on the intersection angle and contact distance.

Using information about contact resistances we estimate the macroscopic conductivity of a composite with nanotube fillers. For this, we used a percolation model based on the Monte Carlo method to construct a nanotube network in a polymer matrix. In this model, we used distributions of contact resistances, obtained from the first-principles calculations for the given angle between nanotubes. Using this Monte Carlo percolation model one can investigate the influence of non-uniformities of a nanotube distribution on macroscopic electrical conductivity. 

In the A section, we will describe the details of molecular dynamics modeling of the atomistic structure of contacts between nanotubes. In the B section, we present the details of first-principles calculations of electron transport for estimates of contact resistance. Finally, in the C section, we present the details of the Monte Carlo percolation model.

\subsection*{A. Preparation of the composite atomic configurations}

Initially, two metallic CNTs with chirality (5,5) were constructed and separated by 6~\AA. The CNTs consisted of 20 periods along the axis, and each one had the total length of 4.92~nm.
The broken bonds at the ends of the CNTs were saturated with Hydrogen atoms.
The distance 6 \AA\ was chosen, because starting with this distance polymer molecules are able to penetrate  the space between CNTs. The three configurations of CNTs junctions were prepared: the first one with parallel CNTs axes (angle between nanotube axes $\varphi=0^\circ$),
the second one with the axes crossing at 45 degrees ($\varphi=45^\circ$), and the third one with perpendicular axes ($\varphi=90^\circ$).

\begin{figure}[b]
\begin{tabular}{ccc}
\includegraphics[width=4cm]{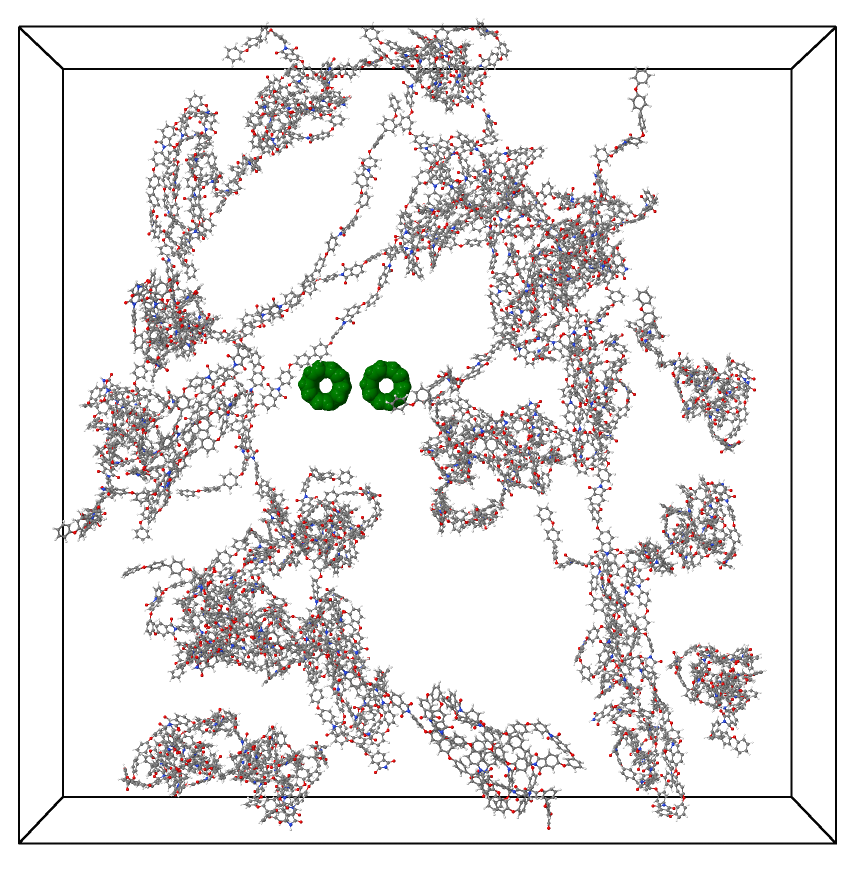}&
\includegraphics[width=4cm]{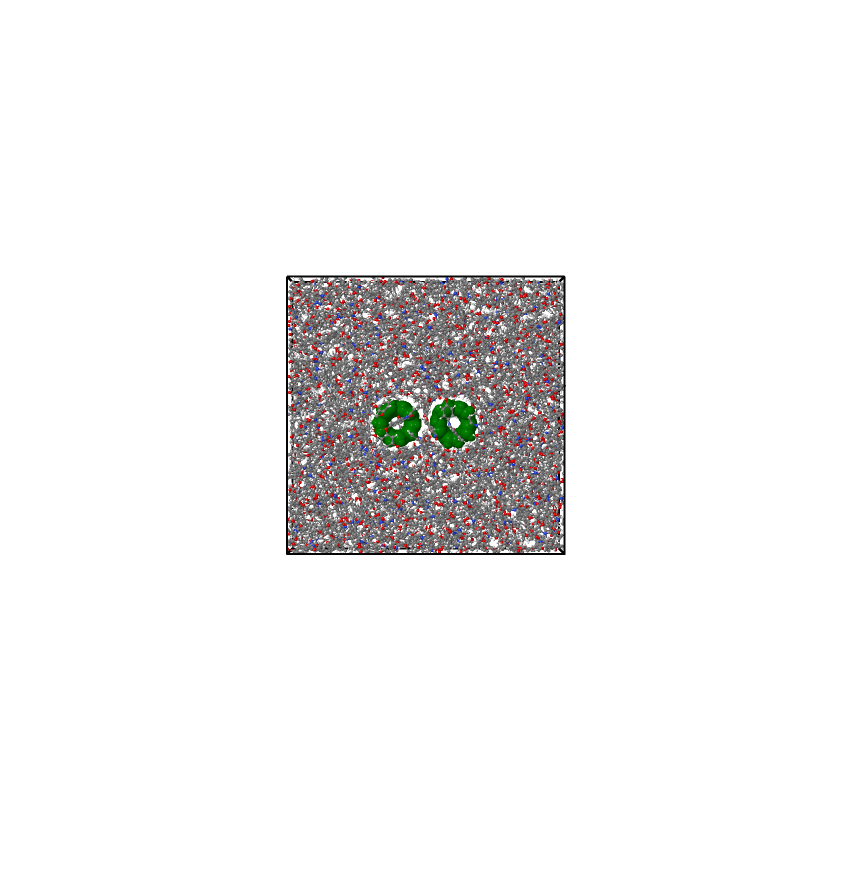}\\
\end{tabular}
\caption{The snapshots of the the nanocomposite system with the parallel orientation of carbon nanotubes at the initial state (left picture) and after the compression procedure (right picture). The black lines represent the periodic simulation cell.} \label{fig_Larin-conf}
\end{figure}

To produce the polymer filled samples, we used 
the procedure  similar to that employed for the simulations of the thermoplastic polyimides and polyimide-based nanocomposites in the previous works
\cite{Larin_RSCADV14, Falkovich_RSCADV14, Nazarychev_EI, Lyulin_MA13, Lyulin_SM14, Nazarychev_MA16, Larin_RSCADV15}. First,  partially coiled R-BAPB chains with the polymerization degree $N_p=8$, which corresponds to the polymer regime onset \cite{Lyulin_MA13, Lyulin_SM14}, were added to the simulation box at random positions avoiding overlapping of polymer chains. This results in the initial configuration of samples with a rather low overall density ($\rho\sim 100~$ kg/m$^3$) (Fig. \ref{fig_Larin-conf}). Then the molecular dynamics simulations were performed to compress the systems generated, equilibrate them and perform production runs.

The molecular dynamics simulations were carried out using Gromacs simulation package \cite{gromacs1, gromacs2}. The atomistic models used to represent both the R-BAPB polyimide and CNTs were parameterized using the Gromos53a6 forcefield \cite{gromos}. Partial charges were calculated using the Hartree-Fock quantum-mechanical method with the  6-31G* basis set, and the Mulliken method was applied to estimate the values of the particle charges from an electron density distribution. As it was shown recently, this combination of the force field and particle charges parameterization method allows one to reproduce qualitatively and quantitatively the thermophysical properties of thermoplastic polyimides \cite{Nazarychev_EI}. The model used in the present work was successfully utilized to study structural, thermophysical and mechanical properties of the R-BAPB polyimide and R-BAPB-based nanocomposites \cite{Larin_RSCADV14, Falkovich_RSCADV14, Nazarychev_EI, Lyulin_MA13, Lyulin_SM14}.

All simulations were performed using the NpT ensemble at temperature $T=600$~K, which is higher than the glass transition temperature of R-BAPB. The temperature and pressure values were maintained using Berendsen thermostat and barostat \cite{Berendses_1, Berendsen_2} with relaxation times $\tau_T= 0.1$~ps and $\tau_p=0.5$~ps respectively. The electrostatic interactions were taken into account using the particle-mesh Ewald summation (PME) method \cite{PME1, PME2}. 

The step-wise compression procedure allows one to obtain dense samples with an overall density close to the experimental polyimide density value ($\rho \approx 1250-1300$~kg/m$^3$), as shown in  Fig. \ref{fig_Larin-conf}. The system pressure $p$ during compression was increased in a step-wise manner up to $p=1000$~bar and decreased then to $p=1$~bar. After compression and equilibration, the production runs were performed to obtain the set  of polymer filled CNT junction configurations.

As the conductance of polymer filled CNT junctions is influenced by the density and structure of a polymer matrix in the nearest vicinity of a contact between CNTs, the relaxation of the overall system density was used as the system equilibration criterion. To estimate the equilibration time, the time dependence of the system density was calculated as well as the density autocorrelation function $C_\rho(t)$:
\begin{equation}
C_\rho(t)=\frac{\langle \rho(0) \rho(t)\rangle}{\langle \rho^2 \rangle},
\end{equation}
where $\rho(t)$ is the density of the system at time $t$ and $\langle \rho^2 \rangle$ is the average density of the system during the simulation.

\begin{figure}[h]
\begin{tabular}{ccc}
\includegraphics[width=7.25cm]{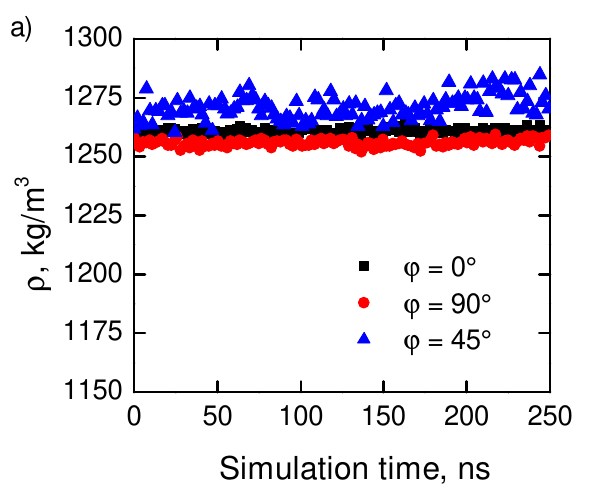}\\
\includegraphics[width=7.25cm]{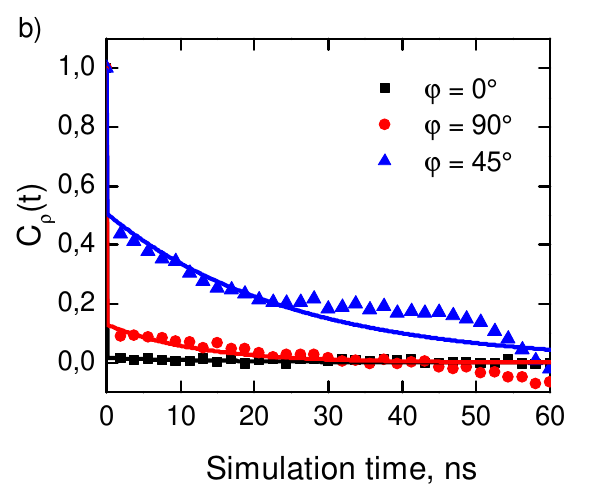}\\
\end{tabular}
\caption{The time dependence of the system density $\rho$ (a) and the density auto-correlation functions $C_\rho(t)$ (b) for the systems with various angles between nanotube axes $\varphi$. The dots correspond to the calculated data. The solid lines correspond to the fitting of $C_\rho(t)$ with the exponential (in case of $\varphi=0^\circ$) or double exponential (in case of of $\varphi=45^\circ$ and $\varphi=90^\circ$) functions.} 
\label{fig_Larin-density}
\end{figure}

As shown in Fig. \ref{fig_Larin-density}a, the system density does not change sufficiently during simulation after the compression procedure. At the same time, the analysis of the density auto-correlation functions shows some difference in the relaxation processes in the systems studied (see Fig. \ref{fig_Larin-density}b). In the case of the system where CNTs were placed parallel to each other ($\varphi=0^\circ$), $C_\rho(t)$ could be approximated by the exponential decay function $C_\rho(t)=\exp(-t/\tau)$ with relaxation time $\tau = 4$~ps. The density relaxation in the systems with crossed CNTs ($\varphi=45^\circ$ and $\varphi=90^\circ$) was found to be slower. For these two systems density the auto-correlation functions could be approximated by a double exponential function $C_\rho(t)=A\exp(-t/\tau_1) + (1-A)\exp(-t/\tau_2)$, and the relaxation times determined using this fitting were $\tau_1=2.7$~ps and $\tau_2=12.2$~ns (for $\varphi=90^\circ$), and $\tau_1=9.5$~ps and $\tau_2=24.6$~ns (in case of $\varphi=45^\circ$).

Nevertheless, the results obtained after the analysis of the system density relaxation allow us to choose the system equilibration time to be 100 ns, which is higher than the longest system density relaxation times determined by the density autocorrelation function analysis. 
The same simulation time was used in our previous works to equilibrate the nanocomposite structure after switching on electrostatic interactions \cite{Larin_RSCADV14, Nazarychev_EI, Larin_RSCADV15}. The equilibration was followed by the 150 ns long production run. To analyze the polymer filled CNT junction conductance, 31 configurations of each simulated system, separated by 5 ns intervals, were taken from the production run trajectory.

After the configurational relaxation is finished, we have to prepare  polymer filled CNT junctions configuration for the first-principles calculations of contact resistance. 
The method we used for the calculations of a  contact resistance is based on the solution of the ballistic electronic transport problem, finding the Volt-Ampere characteristic $I(V)$ of a device  and
deriving the contact resistance from the linear part of $I(V)$ corresponding to the low voltages. For this purpose, we employed the Green's function method for 
solving the scattering problem and the Landauer-Buttiker approach to find the current through a scattering region coupled to two semi-infinite leads, as described in \cite{Datta}.
Specific details of how these techniques are applied in the case of crossed CNTs can be found in \cite{comp_no_pol}.

\subsection*{B. The first-principles calculations of the contact resistance of CNTs junctions filled with polymer} \label{sec_meth_fp}

For the preparation of a device for the electronic transport calculations, we first form that part of the device which consists of the atoms belonging to the CNTs used in
the CNTs+polymer relaxation. Regions with the same geometry as in \cite{comp_no_pol}
are cut from the initial 20-period long CNTs, and the rest of the atoms belonging to the CNTs are discarded. This is done to make possible a direct comparison of the results obtained
for the polymer filled CNTs junctions with the results for CNTs junctions without polymer reported in  \cite{comp_no_pol} for the same separation of CNTs equal to 6 \AA. 

Note that
the CNTs parts of the scattering device contain atoms  shifted from their positions in ideal CNTs due to the influence of the adjacent polymer molecules, and these shifts are time-dependent
as a result of thermal fluctuations. 

The cut regions contain two fragments of CNTs each 9 periods long, and in the case of the CNTs parallel configuration, the CNTs overlap by 7 periods.
In the nonparallel configurations, one of the CNTs is rotated around the axis perpendicular to the CNTs axes in the parallel configuration and passing through the
geometrical center of a device in the parallel configuration. 

\begin{figure}[b]
\begin{tabular}{ccc}
\includegraphics[width=2.25cm]{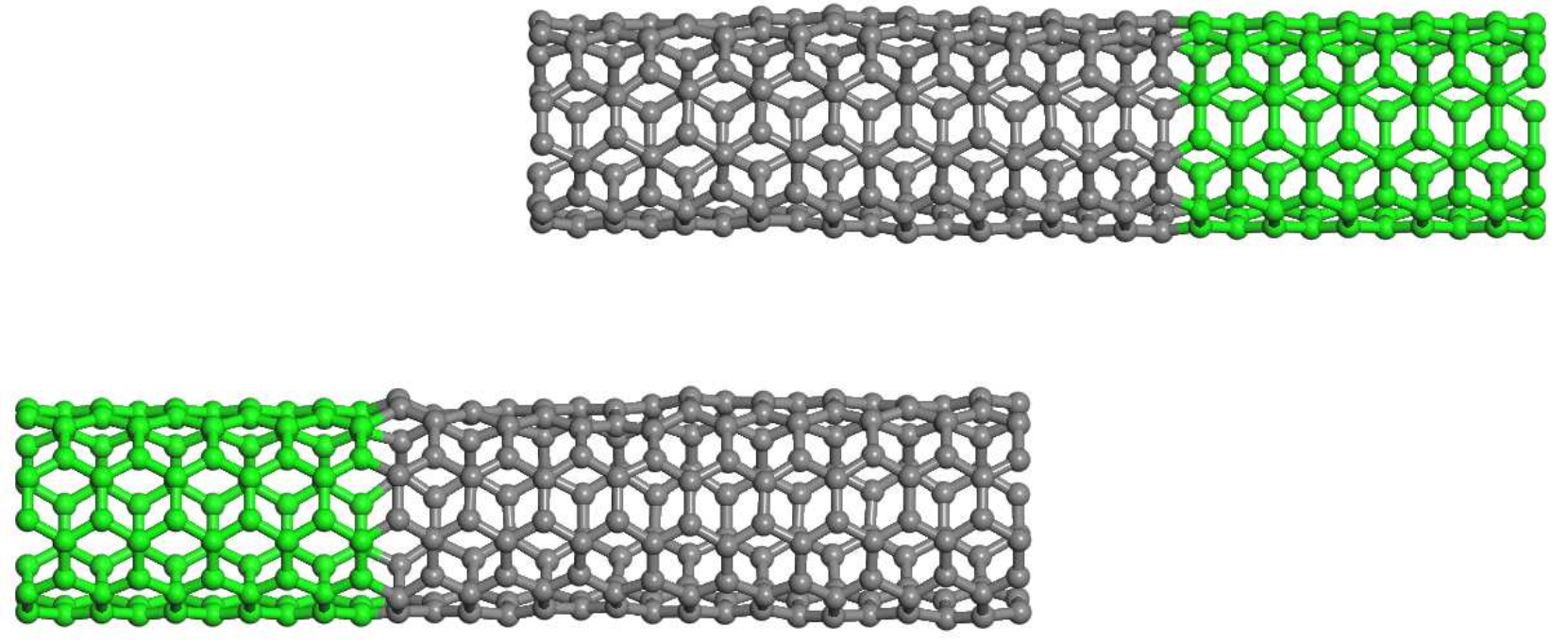}&
\includegraphics[width=2.25cm]{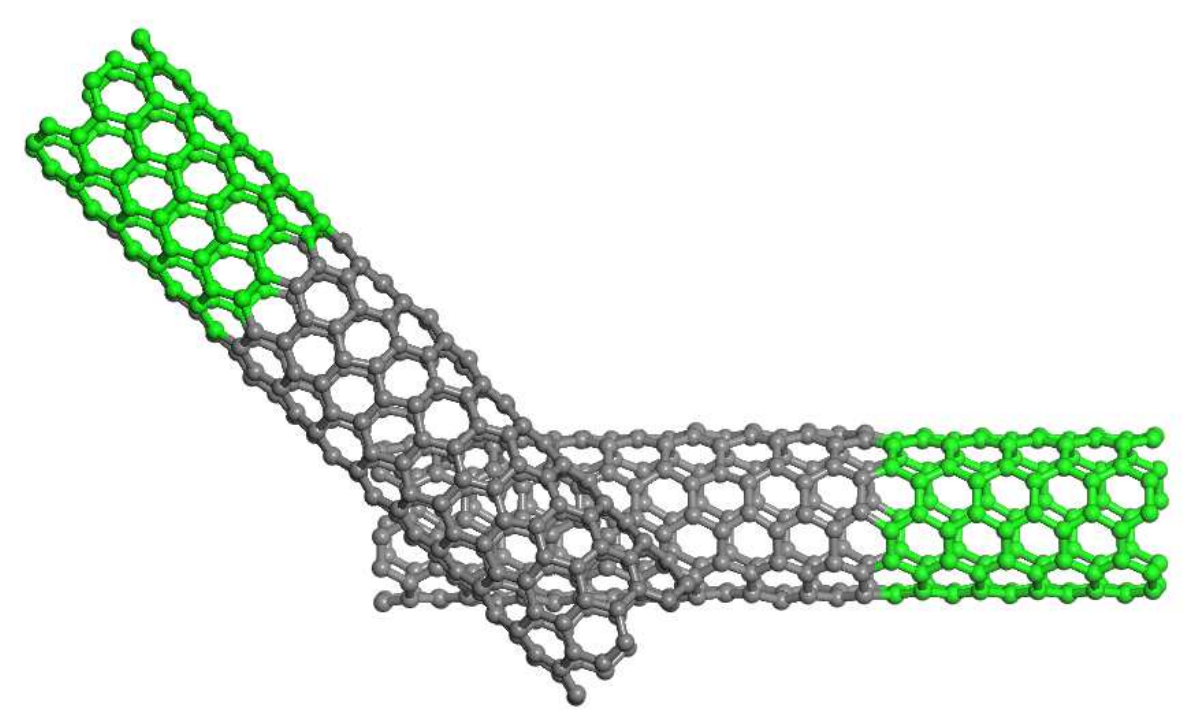}&
\includegraphics[width=2.25cm]{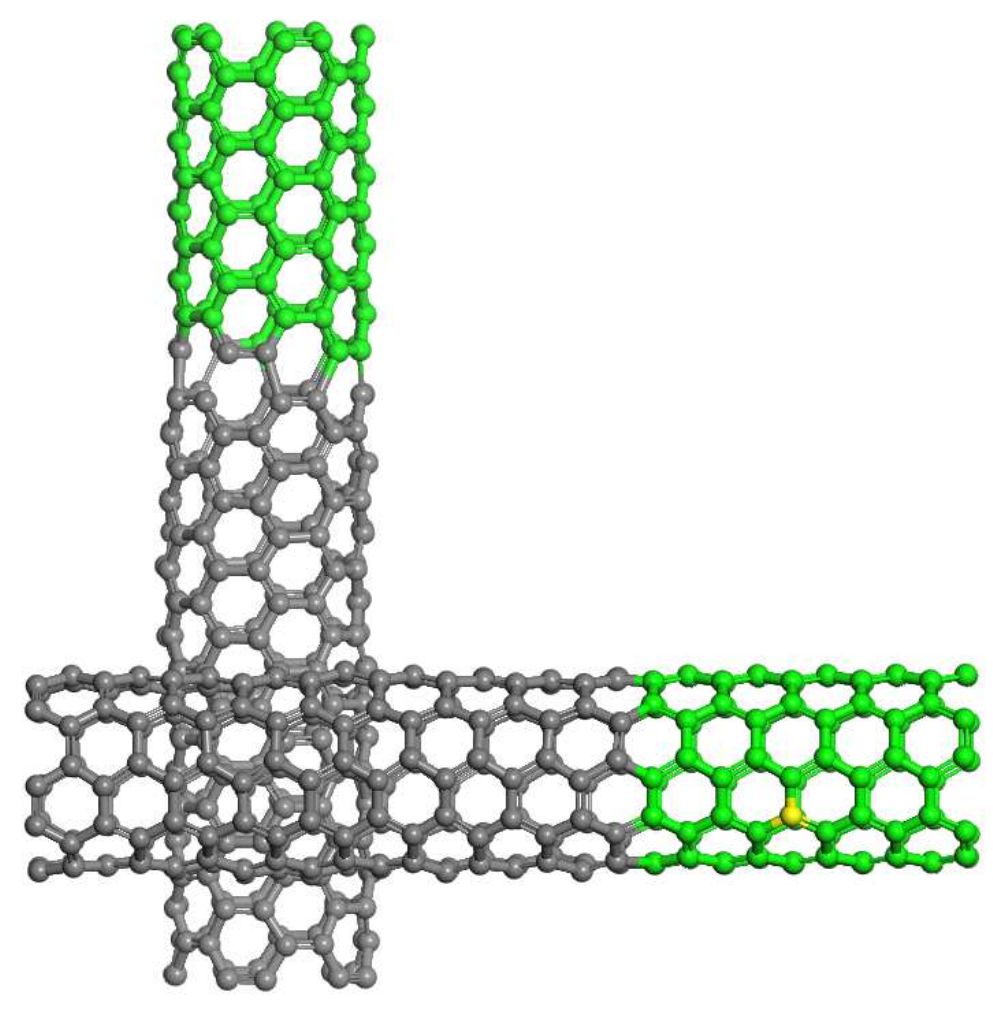}\\
\end{tabular}
\caption{The CNTs parts of the junctions. Left: the parallel configuration, center: CNTs axes are crossing at 45 degrees,
right: the perpendicular configuration. The leads atoms are colored by green. } \label{fig_CNT_conf}
\end{figure}

After the construction of the CNTs part of the scattering region, we attach to it leads that consist of 5 period long fragments of an ideal CNT. The CNTs parts of the scattering regions with the attached leads for the three considered configurations are shown in Fig. \ref{fig_CNT_conf}.

After the preparation of the CNT parts of the junctions, we still have 17766 atoms in a device. A system with such a large number of atoms cannot be treated by fully first-principles
atomistic methods. On the other hand, keeping all those atoms for  a precision calculation of the contact resistance of polymer filled CNTs junctions is not necessary,
as only those polymer atoms which are close enough to a CNT will serve as tunneling bridges and give a contribution to the junctions conductivity.  Thus,
for the calculations of the contact resistance, only those atoms were kept  which are closer to the CNTs than a certain distance $d$. It has been established by numerical experiments
that if the value of $d$ is taken equal to the CNTs separation $d=6$ \AA\  this is quite sufficient, and taking into account more distant atoms does not change the contact resistance significantly.

The procedure of sorting the polymer atoms is as follows. In our molecular dynamics simulations we used 27 separate polymer molecules,  consisting of 8 monomers each.
If at least one  of the atoms of a polymer molecule was closer to the CNTs part of a junction than $d=6$ \AA, the whole molecule was kept for  a while, and  discarded 
otherwise. Having applied this first part of the procedure, we kept 4 polymer molecules for the parallel configuration, 8 molecules for the perpendicular configuration and
11 molecules for the 45 degrees configuration. 

Then we looked at the monomers of each molecule that survived the first round of selection. The same procedure was applied
to monomers as the one used earlier for molecules: if at least one  of the monomer atoms was closer to the CNTs part of a junction than $d=6$ \AA, 
the whole monomer was kept for  a while, and 
discarded  otherwise. 

\begin{figure}[b]
\begin{tabular}{ccc}
\includegraphics[width=2.25cm]{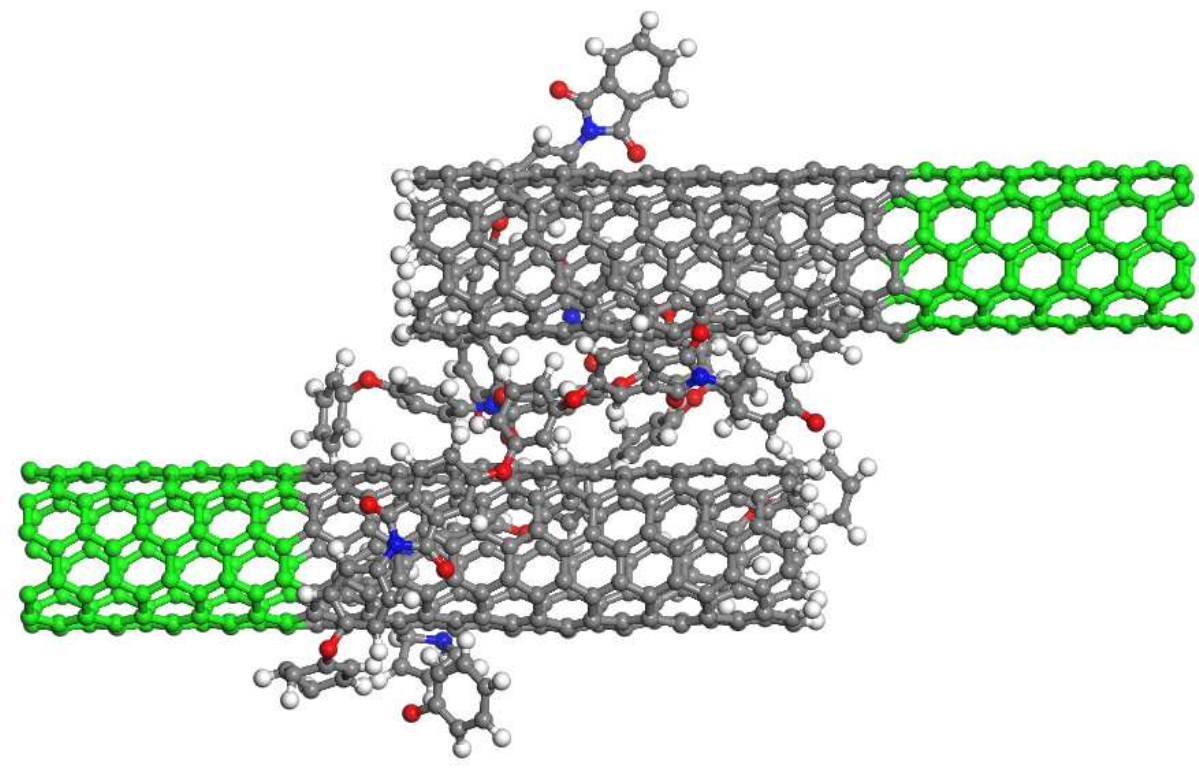}&
\includegraphics[width=2.25cm]{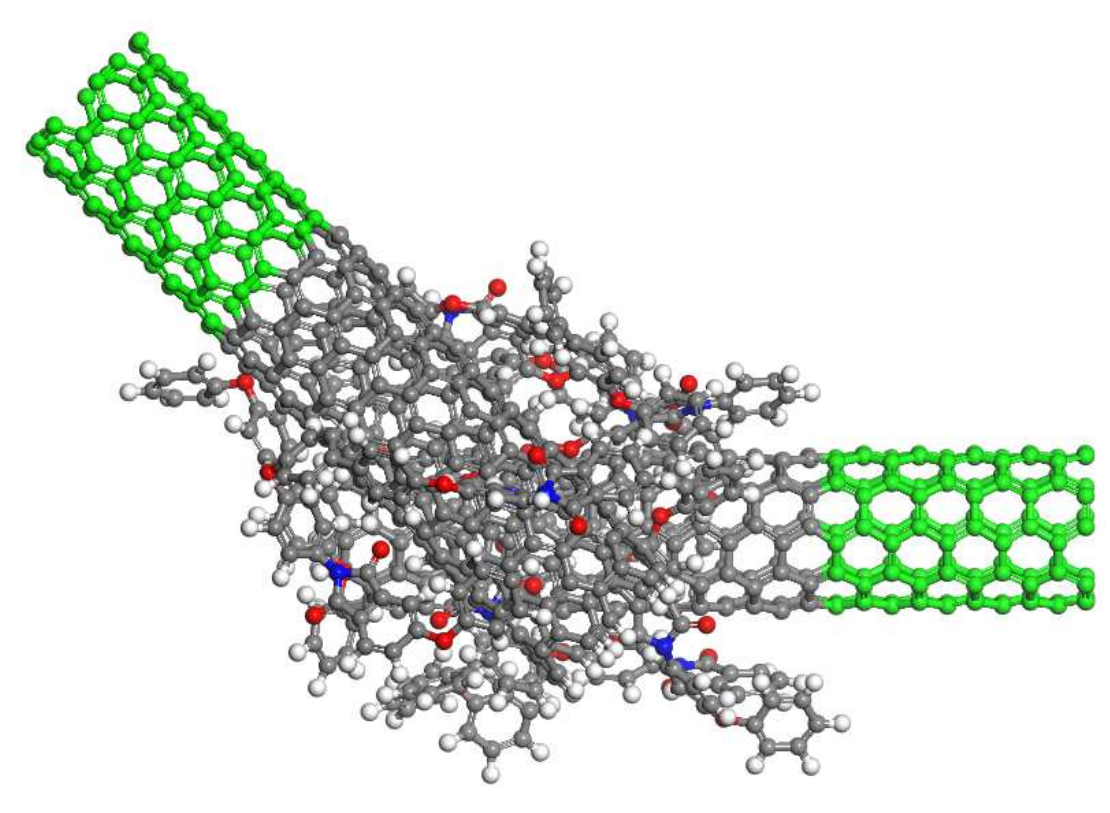}&
\includegraphics[width=2.25cm]{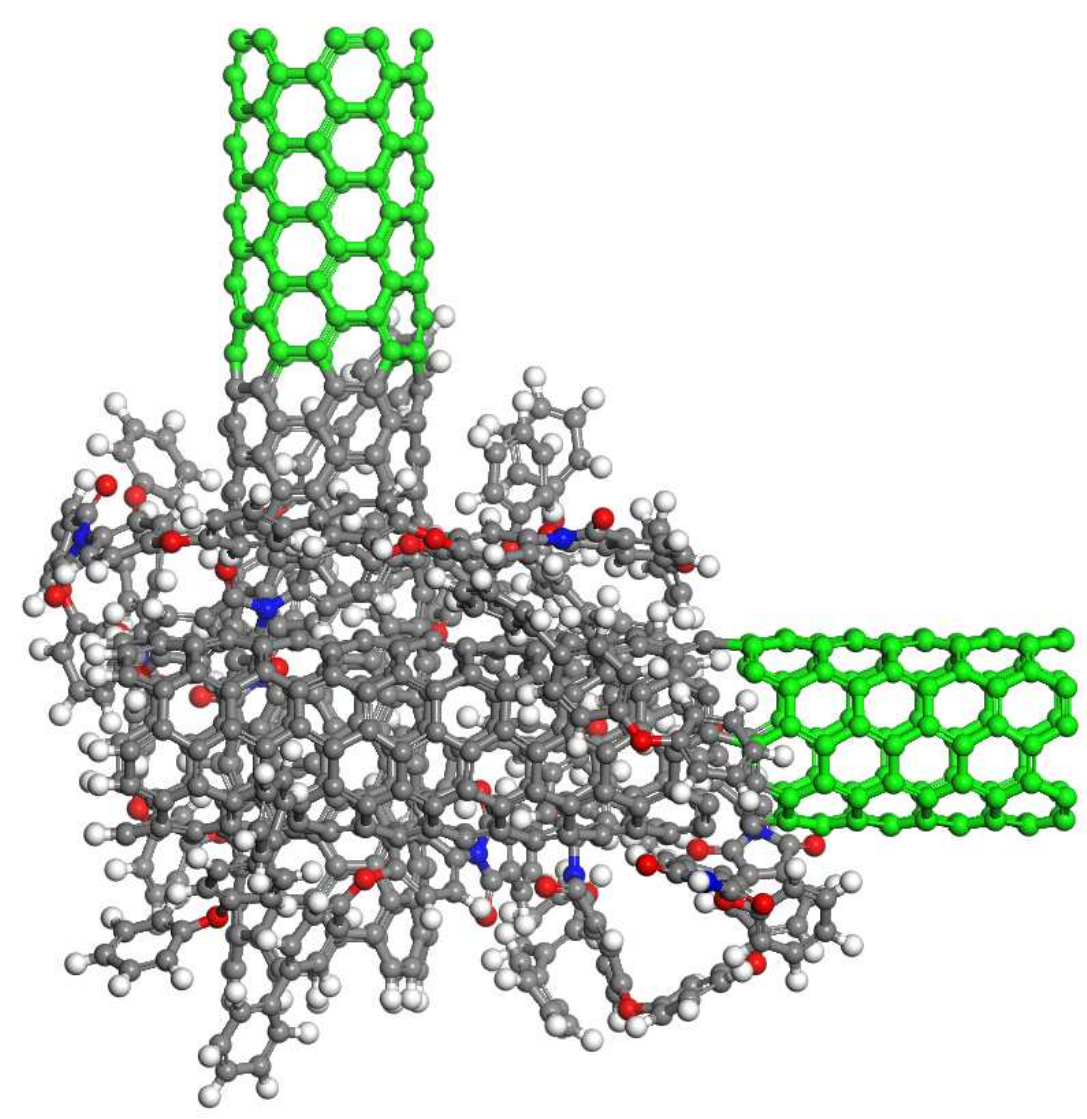}\\
\end{tabular}
\caption{The atomic configurations for the first-principles calculations of the polymer filled CNTs junctions contact resistance. The configurations are for the first time steps in the corresponding series.
Left: the parallel configuration, center: CNTs axes are crossed at 45 degrees,
right: the perpendicular configuration. The Carbon atoms are gray, the Nitrogen atoms are blue, the Oxygen atoms are red, and the Hydrogen atoms are light gray. The leads atoms are colored by green. } 
\label{fig_junction_conf}
\end{figure}

After the second round of selection with monomers was over, we dealt in the same manner with the individual residues comprising a monomer. The broken
bonds that appeared in the second and third stages were saturated with hydrogen atoms. The described procedure resulted in the following numbers of atoms in the whole
device, including the central scattering region and the leads: 881 for the parallel configuration, 1150 for the perpendicular configuration, and 1074 for the 45 degrees configuration.
The atomic configurations obtained using the described procedure for the first time steps in the corresponding series are presented in Fig. \ref{fig_junction_conf}.

A fully ab-initio method for electronic structure investigations utilizing a localized pseudo-atomic basis set, as described in \cite{Ozaki_DFT} and implemented in \cite{OpenMX}, 
was used for the calculations of the electronic structures of the whole device and the leads.
We used basis set s2p2d1, the Pseudo Atomic Orbitals (PAO) cutoff radius equal to 6.0 a. u., and the cutoff energy of 150 Ry.
The pseudo-potentials generated according to Morrison, Bylander, and Kleinman scheme \cite{pseudopot} were used. For the 
density functional  calculations, the exchange-correlation functional was used in PBE96 form \cite{PBE96}.

Using the electronic structures of the whole device and the leads we calculated energy the dependent transmission function
 through the device. Then the dependence  $I(V)$ of the current $I$ on the voltage $V$ between the leads   was determined
with the Green's function approach as described in detail in \cite{Datta}. Finally, the Landauer-Buttiker approach was used to find the current through a polymer filled CNTs junction.

Solving a scattering problem for a nano-device at arbitrary voltages is a computationally very complex task since it requires achieving self-consistency for both
electron density and induced electrostatic potential simultaneously. Fortunately, for  contact resistance calculations one can take
advantage of the fact  that the required voltages are very low. 

According to the experimental evidence,  the size of a nanocomposite specimen used in conductivity experiments is about 10 mm \cite{exp_CNT_length_1}, and the typical voltages applied across such specimen do not exceed 100 V \cite{CNT_book}. Then for the size of a central scattering region about 1 nm, the voltage drop is about 10$^{-5}$ V, which is well within the range where the simplified approach is applicable.

The question of the modeling of quantum transport in the limit of low voltages was discussed in detail in \cite{comp_no_pol}, where it was demonstrated that in the case of moderate voltages between leads,
the scattering probability $T(E)$ is not sensitive to the details of the electrostatic potential distribution $V({\mathbf r})$ in the central scattering region, and
some physically reasonable approximation may be chosen for $V({\mathbf r})$. 

This is due to the fact that the difference of the Fermi functions
$f(\varepsilon - \mu_L)$ and  $f(\varepsilon - \mu_R)$ 
for the left and right leads with corresponding the chemical potentials $\mu_L$ and $\mu_R$, present in the original Landauer-Buttiker formula:
\begin{equation}
I=\frac{2e}{h} \int T(\varepsilon) \left (f(\varepsilon - \mu_L) - f(\varepsilon - \mu_R) \right )d\varepsilon ,
\label{LB_formula}
\end{equation}
where $e$ is the elementary charge, $h$ is the  Planck constant, $\varepsilon$ is the electron energy and  $T(\varepsilon)$ 
is the energy dependent transmission probability,
 is reduced in this case to a very narrow and sharp peak centered at the Fermi level
of the device. 

\begin{figure}[b]
\includegraphics[width=8cm]{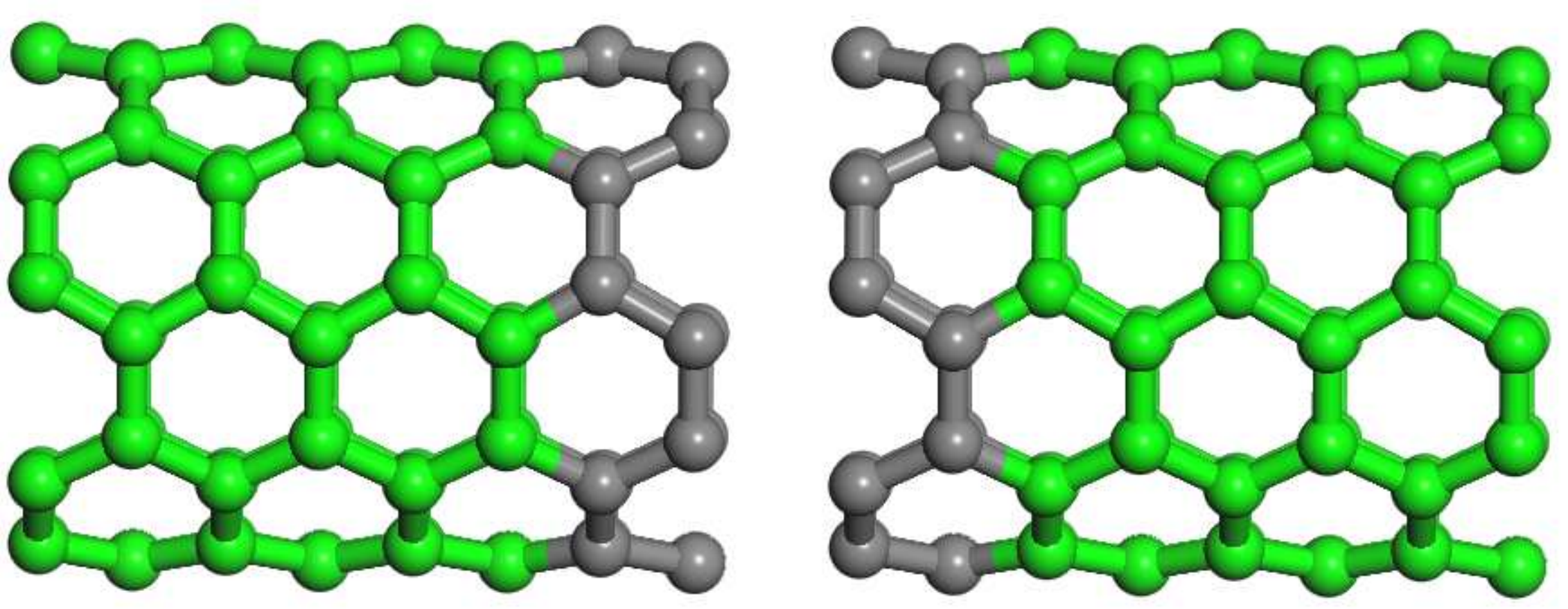}
\caption{The atomic configuration used in this work to validate the method of calculating quantum transport by comparing it againts the consistent NEGF approach.} \label{fig_NEGF_compare}
\end{figure}

In addition to the analysis performed in \cite{comp_no_pol}, in this article, to verify the accuracy of our  approach, we made contact resistance calculations for a simple test CNT junction in a coaxial configuration, using both the simplified method we suggest and the full NEGF method, where not only electron charge density but the electric potential was converged as well. The interlead voltage used in those test calculations was
set to $10^{-4}$ V, and the gap between the CNTs tips was 0.94 \AA. The atomic configurations for the test calculations are presented in Fig. \ref{fig_NEGF_compare}. The consistent NEGF calculations produced 1.71~$\cdot 10^{-5}$ S for the conductance of the junctions shown in Fig. \ref{fig_NEGF_compare}, while modeling without searching for convergence of potential  yielded 1.72 $\cdot 10^{-5}$ S.  

Thus, in our case, a  very complex task of finding the $I(V)$ characteristic of a nano-device can be significantly simplified without the loss
of precision. For the $I(V)$ calculations in the current paper we employed the abrupt potential model introduced in \cite{comp_no_pol}: the potentials $V_L$ for the
left lead and $V_R$ for the right lead were set and were used for all atoms of the corresponding CNT to which that lead belonged. As for the polymer atoms, both $V_L$ and
$V_R$ can be safely used for them, and at the considered voltages, adopting these two options, as we have checked by direct calculations, leads to the differences in
current not exceeding 0.1\%.

\subsection*{C. The percolation model} \label{sec_percolation}
Determination of the conductivity of a polymer-CNT system can be implemented in 2 stages. First, a percolation cluster is formed, and  the second stage implies solving the matrix problem for a random resistor circuit (network). 

At the first stage, the modeling area -- a cube of the linear size  $L$ -- is filled with CNTs. For this task, permeable capsules (cylinders with hemispheres at the ends) with a fixed length and diameter were chosen as filling objects corresponding to CNTs. The cube is filled by the successive addition of CNTs until a fixed bulk density of CNTs   
\begin{equation}
\eta = \frac{((4/3)\pi R^3 + \pi  R^2 h)N}{L^3}  \label{cnt_vol_frac}
\end{equation}
in the cube is reached, where $R$ is the radius of the cylinder and hemisphere,  $h$ is the height of the cylinder,  and $N$ is the number of CNTs in the cube.  The percolation problem for permeable capsules was previously solved in \cite{Xu_16}, and in \cite{Schilling_15} capsules with a semipermeable shell were considered.

In the percolation problem, we use periodic boundary conditions as, for example,  in \cite{Bao_2011}. 
We use the method of finding a percolation threshold based on the Newman and Ziff algorithm \cite{Newman_2001}, where  the identification of a percolation cluster is made at the stage of its formation.

When a percolation cluster is formed, the obtained CNT configuration is transformed into a resistor circuit (2nd stage). 

The contributions to a conductance matrix resulting from inner resistance of CNTs and the junctions tunneling resistance are usually discussed in connection with constructing conductivity percolation algorithms.  Direct measurements of CNTs resistance per unit length are available. In \cite{Chiodarelli_11}, the inner resistance of CNTs is estimated as $15 \cdot 10^3$ $\Omega/\mu$m. The results of \cite{Ebbesen_96} give specific CNTs resistance in the range $(12-86) \cdot 10^3$ $\Omega/\mu$m. Taking into account that the characteristic CNTs lengths in nanocomposites are about several $\mu$m \cite{exp_CNT_length_1,CNT_book}, this results in the inner CNTs resistance approximately $10^4-10^5$ $\Omega$, which is at least one order of magnitude less than the tunneling resistance obtained in this work. The specific results on tunneling resistance will be discussed below in section \ref{RND}. Thus, in our percolation model, the inner resistance of CNTs is neglected, and only tunneling resistance of CNTs junctions is taken into account. This can significantly reduce the requirements for computational time. 

When contact resistance is determined only by tunneling, the principle of compiling the matrix for the percolation problem will be as follows. 
First, the matrix (N, N) is compiled from the bonds of percolation elements, where N is the number of CNTs participating in percolation. 
Then this matrix is transformed into the conductivity matrix according to the second Kirchhoff law 
\begin{equation}
\sum_j G_{ij} (V_i-V_j) = 0 \label {second_Kirchoff}
\end{equation}
- the sum of currents for all internal elements of a percolation network is zero -- where $G_{ij}$ are the elements of  the  conductance matrix ${\mathbf G}$ and $V_i$ is the component of the voltage vector ${\mathbf V}$  corresponding to the $i$-th contact point in a percolation network \cite{Kirkpatrick_73}. 
The voltages on the left and right borders of a simulation volume are set to $V_L=1$ V and $V_R=0$ V, respectively. 

Now finding the conductivity of the system is reduced to the problem  $\mathbf{GV}=\mathbf{I}$, where
$\mathbf{I}$ is the vector of the currents between the contact points.  

The dimension of the matrix problem can be further reduced to 
(N-2, N-2) by excluding boundary elements. 

After solving the equation (\ref{second_Kirchoff}), with the elements of the $\mathbf{G}$ matrix obtained by the first-principles calculations, we obtain the voltage vector for all internal elements. 
Then, knowing this vector, we sum up  all the currents on each of the boundaries. The currents on the left $I_L$ and right $I_R$ 
boundaries of a simulation volume are equal in magnitude and opposite in sign $I_L= - I_R$. Knowing these currents, we determine the conductance of the simulation system as 
$G=|I_L|/(V_L-V_R) = |I_R|/(V_L-V_R)$. 
And then the conductivity  of the composite is calculated  as $\sigma=GL/S$, where $L$ is the distance between the faces of a simulation volume where voltage is applied, and $S$ is  the area of that kind of face. In our case, for the simulation volume of a cubic shape, $S=L^2$, and  $\sigma=G/L$.

To calculate the conductivity, the following system parameters were selected: the length of a nanotube is $l=3$ $\mu$m, the diameter of a CNT is $D=30$ nm, the aspect ratio $l/D=100$, and the size of the system is 4 $\mu$m. The same values were used in \cite{Yu_2010}. We adopted those values to test our realization of the percolation algorithm against the previously obtained results \cite{Yu_2010}.
Then for the given parameters for each fixed tube density, the Monte Carlo method (100 implementations of various configurations of CNT networks) was  used to calculate the system conductivity. 

The quality of CNTs dispersion is one of the key factors that affect the properties of nanocomposites, and a lot of efforts is taken
to achieve a homogeneous distribution of fillers \cite{Mital_14}. In this work, we take into consideration the effect of inhomogeneity of a CNTs distribution on composite conductivity.
The spatial density of nanotubes  $\rho_{CNT}$, in this case, has one peak with a Gaussian  distribution:
\begin{equation} 
\rho_{CNT}=\rho_0 \cdot exp(-(\mathbf {r-r_0})^2/\rho_\sigma^2), \label {agglo}
\end{equation}
where  ${\mathbf r_0}$ coincides with the geometrical center of a simulation volume, and $\rho_\sigma = L/12$.  The value of the $\rho_0$ parameter is chosen so that the  CNTs volume fraction in the inhomogeneous case is the same as in the homogeneous distribution.

\section*{Results and discussion}\label{RND}
\begin{figure}[b]
\begin{tabular}{lr}
\includegraphics[width=3.5cm]{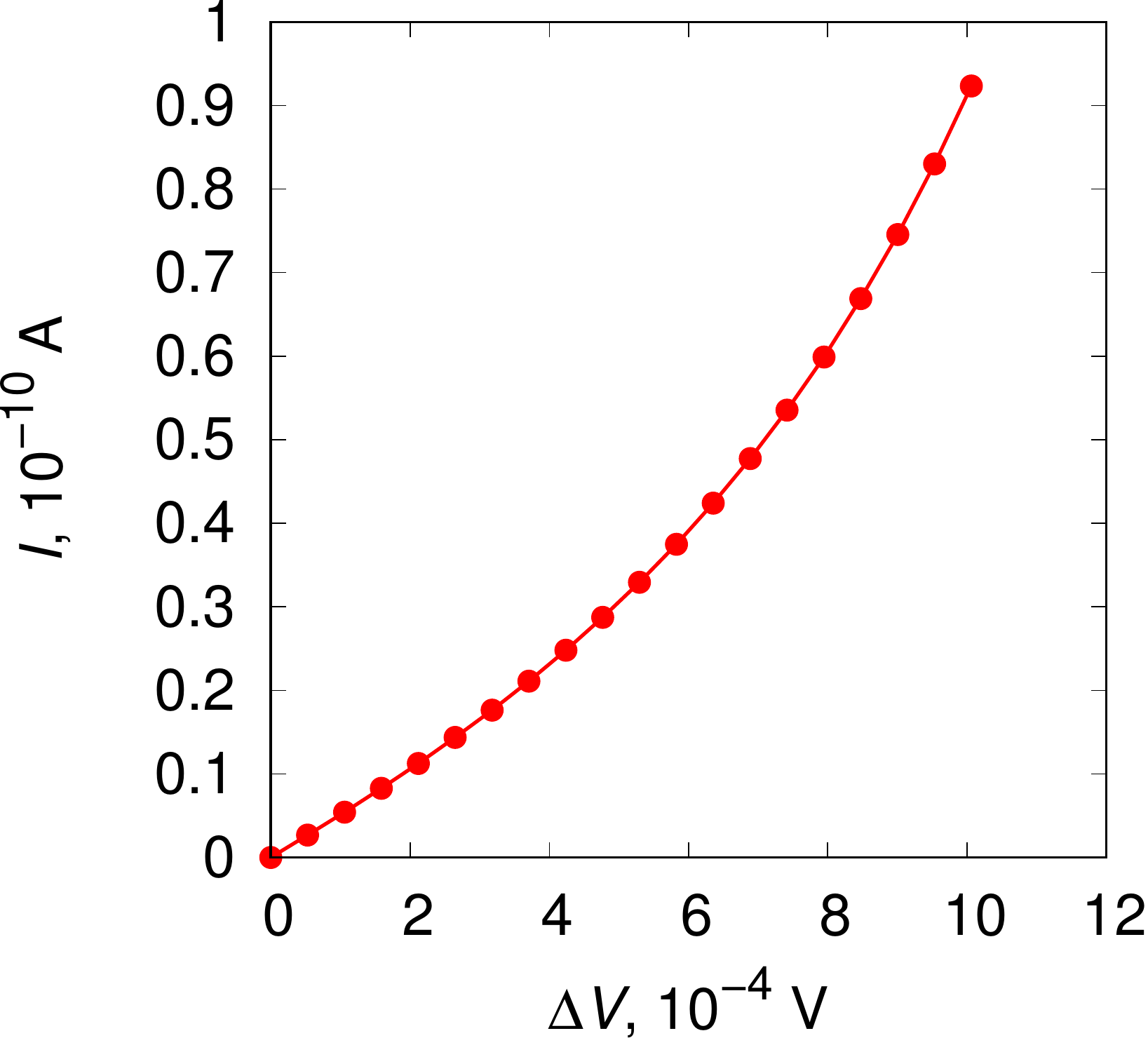} &
\includegraphics[width=3.5cm]{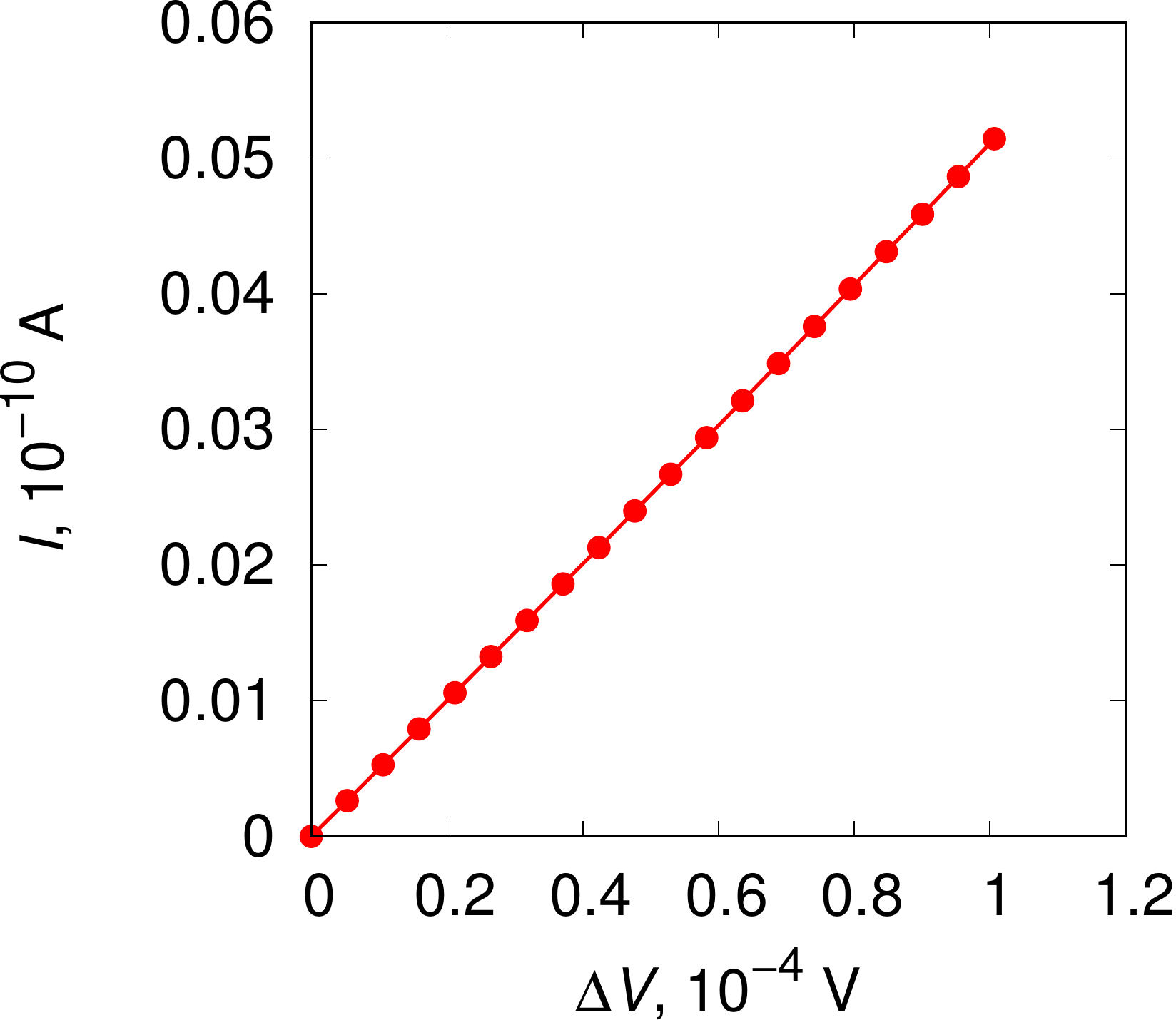}
\end{tabular}
\caption{The Volt-Ampere characteristic for the polymer filled CNTs junction corresponding to the first time step in the series for the parallel configuration.
Left frame: maximum inter-lead voltage is $10^{-3}$ V, right frame:  $10^{-4}$ V. The circles correspond to the results of calculations,
the lines are guides for an eye.} 
\label{fig_I_V}
\end{figure}

To find the contact resistance of polymer filled CNTs junctions one first needs to find their Volt-Ampere characteristics $I(V)$ and to determine the voltage range where
$I(V)$ is linear and is not sensitive to the specific distribution of an electrostatic potential in the scattering region. In Fig. \ref{fig_I_V} the $I(V)$ plot
for the first time step in the atomic geometry series for the parallel configuration is shown.

It is clearly seen from Fig. \ref{fig_I_V} that up to about $10^{-4}$ V the $I(V)$ characteristic is linear, and after that value, it starts to deviate from a
simple linear dependence. Thus, for the calculations of a contact resistance $R$ and its inverse, a junction conductivity $G$,
we used the electrical current values obtained for the inter-lead voltage equal to $10^{-4}$ V. Note, that according to our estimates in section \ref{sec_meth_fp}, a characteristic voltage drop on the length of a CNTs tunneling junction is about $10^{-5}$ V which is well within the region where the linear $I(V)$ is observed.

\begin{figure}
\includegraphics[width=8cm]{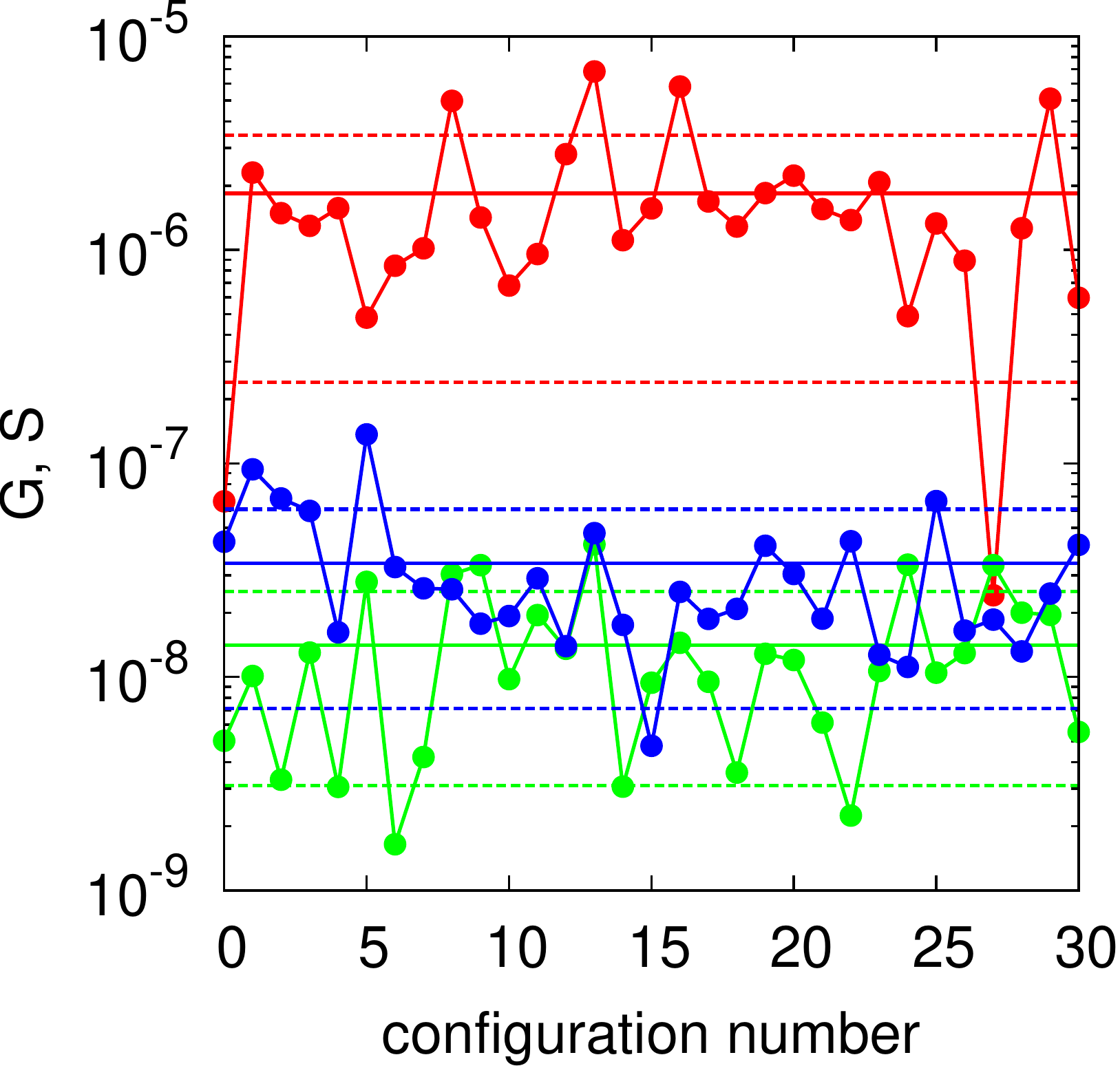}
\caption{The time dependence of the polymer filled junctions conductance $G$ in S. Red color correspond to the parallel configuration, the green lines -- to the perpendicular configuration
and the blue lines -- the 45 degrees configuration. The results of the calculations are shown by circles, the saw-tooth lines serve as a guide for an eye.
The straight solid lines designate the mean values of conductance $\langle G \rangle$, and the dashed ones -- $\langle G \rangle \pm G_\sigma$.} 
\label{fig_G_t}
\end{figure}

The time dependences of the junctions conductances for the three considered configurations are presented in Fig. \ref{fig_G_t}. One might expect that the 
shifts of both CNTs atoms and polymer atoms in the central scattering region due to thermal fluctuations would lead to fluctuations of junctions conductances $G$, but 
quantitative characteristics of this phenomenon such as minimum $G_{min}$, maximum $G_{max}$, mean values $\langle G \rangle$ and a standard deviation $G_\sigma$ can only be captured by highly precise fully atomistic 
first-principles methods, like those employed in the current paper. The resulting fluctuations of conductance are very high. For the parallel CNTs configuration
the minimum value, $G_{min} = 2.4 \cdot 10^{-8}$ S, and the maximum value, $G_{max} = 6.8 \cdot 10^{-6}$ S, differ by more than two orders of magnitude, for the 45 degrees and perpendicular
configurations the corresponding ratios are about 30. The same strong variations of conductance over time were reported in \cite{Penazzi_2013} for the coaxial
CNTs configuration, where the results were obtained using a semi-empirical tight-binding approximation. Thus, it is obvious that for the precise determination
of a conductance  of polymer filled CNTs junctions one needs to use fully atomistic approaches, and phenomenological methods taking atomic configurations into account 
on the average are not reliable.

\begin{table*}[h]
 \centering
\caption{The results of statistical analysis of the CNTs junctions conductances in S, for CNTs separation equal to 6 \AA \ and different values of the CNTs crossing angles $\varphi$, 
without polymer from \cite{comp_no_pol}, and with polyimide R-BAPB filling obtained in the current paper.
}
\begin{tabular}{cccccc}
\textrm{$\varphi$}&
\textrm{no polymer}&
\textrm{}&
\textrm{polymer present}&
\textrm{}&
\textrm{}\\
\textrm{}&
\textrm{the results of \cite{comp_no_pol}}&
\textrm{$G_{min}$}&
\textrm{$G_{max}$}&
\textrm{$\langle G \rangle$}&
\textrm{$G_\sigma$}\\
\hline
& & & & &\\  
0 &  3.6$\cdot10^{-13}$ &2.4$\cdot10^{-8}$ &6.8$\cdot10^{-6}$ & 1.8$\cdot10^{-6}$ & 1.6$\cdot10^{-6}$\\
& & & & &  \\
\hline
& & & & &\\ 
0.2$\pi$        & 1.4$\cdot10^{-14}$ & --- & --- & --- &\\
& & & & &\\ 
\hline
& & & & &\\ 
0.25$\pi$  & --- &4.8$\cdot10^{-9}$ &1.4$\cdot10^{-7}$ &3.4$\cdot10^{-8}$ &2.7$\cdot10^{-8}$\\
& & & & &\\ 
\hline
& & & & &\\ 
0.3$\pi$        & 1.2$\cdot10^{-14}$ & --- & --- & --- &\\
& & & & &\\ 
\hline
& & & & &\\ 
0.5$\pi$        & 4.2$\cdot10^{-14}$ &2.2$\cdot10^{-9}$ &4.3$\cdot10^{-8}$ & 1.4$\cdot10^{-8}$ & 1.1$\cdot10^{-8}$\\
& & & & &\\ 
\end{tabular}
\label{tab_cond}
\end{table*}

To assign the tunneling resistance to a polymer filled CNT junction the following algorithm was used. First, for each junction that had to be used in the percolation algorithm, a uniformly distributed random number $\varphi$ in the range [0, $\pi/2$] was generated. The value of the intersection angle  for that junction  was assigned to the obtained random number. The mean values and standard deviations for CNTs tunneling resistances and conductances calculated for the different atomic configurations corresponding to the different time steps are known for $\varphi = 0, \pi/4$, and $\pi/2$. Analyzing figure 4 of \cite{comp_no_pol}, one can see that though an angle dependence of current and hence conductivity is a rather complex function, in the first approximation one can adopt a roughly  piece-wise linear character for this function with the minimum located at $\varphi=0.25\pi$. Thus the logarithm of the mean value of conductance $\mu_\varphi$ for the generated $\varphi$ was set by linear interpolation between the logarithms of the mean values of conductances for $\varphi=0$ and $\varphi=\pi/4$ or $\varphi=\pi/4$ and $\varphi=\pi/2$ presented in Table \ref{tab_cond}. The same algorithm was applied to finding the standard deviation values $\sigma_\varphi$ for the generated $\varphi$. After the statistical parameters for the generated $\varphi$ are estimated, the  conductivity of the junction is set to a random number generated using the normal distribution with the parameters $\mu_\varphi$ and $\sigma_\varphi$.

In \cite{comp_no_pol}, the conductances were reported for the CNTs junctions with almost the same geometry as the CNTs parts of the devices considered in the current paper. The only difference 
between the configurations is that in this work the carbon atoms belonging to the CNTs part of the central scattering region are shifted somewhat from their equilibrium positions
due to the interaction with polymer. The maximum values of those shifts along the $x$, $y$, and $z$ coordinates lie in the range $0.2-0.5$ \AA. This gives us the possibility to directly compare  the current results to the data 
from \cite{comp_no_pol} and,
thus, elucidate the influence of polymer filling on the junctions conductance. The corresponding data and the results of a basic statistical
analysis for the case of the polymer filled junctions are provided in Tab. \ref{tab_cond}.

First, as was expected, filling CNTs junctions with polymer creates carrier tunneling paths and increases junctions conductance by 6-7 orders of magnitude.
Second, it is evident that the CNTs axes crossing angle is crucial for the junctions conductivity when polymer is present as it was the case without
polymer  \cite{comp_no_pol}. At the same time, the sharp dependence of polymer filled junctions conductance on the CNTs crossing angle is somewhat different
from the analogous dependence for junctions without polymers. While in the latter case this dependence is sharply non-monotonous, with a pronounced minimum
at the angles around $0.25\pi$, in the former case there is a significant difference between the conductance values for the parallel and nonparallel configurations, but
the configurations with the angle $\varphi$ between CNTs angles equal to $0.25\pi$ and $0.5\pi$ have very close conductances, and their mean values averaged over
time  $\langle G \rangle_{45}$ and $\langle G \rangle_{per}$  lie within the
ranges $\langle G \rangle \pm G_\sigma$ of each other. Moreover, in contrast to the geometries without polymer, for the polymer filled CNTs junctions  $\langle G \rangle_{per}$ is lower than  $\langle G \rangle_{45}$ by a factor of 2.4.

Note also that for the parallel configuration, the polymer influence on the junction conductance is more pronounced than for the nonparallel ones.
For the parallel configuration, adding polymer to a junction of CNTs separated by 6.0 \AA\  with initial conductance of  3.6$\cdot 10^{-13}$ S 
produces conductance mean value equal to 1.8$\cdot 10^{-6}$ S. This gives the factor 0.5$\cdot 10^7$; the value of the analogous factor for the perpendicular configuration is 0.33$\cdot 10^6$.

The probable reason for the more effective conductance increase, when polymer is added,  for the configurations  with smaller angles between CNTs axes, is
that the smaller is an intersection angle, the larger is the overlap area between CNTs where polymer can penetrate and, thus, create tunneling bridges.
The higher fluctuation of conductance over time for the parallel configuration can be explained by the same reason: a larger CNTs overlap area gives
more freedom for polymer atoms  to adjust their positions.

\begin{figure}
\includegraphics[width=8cm]{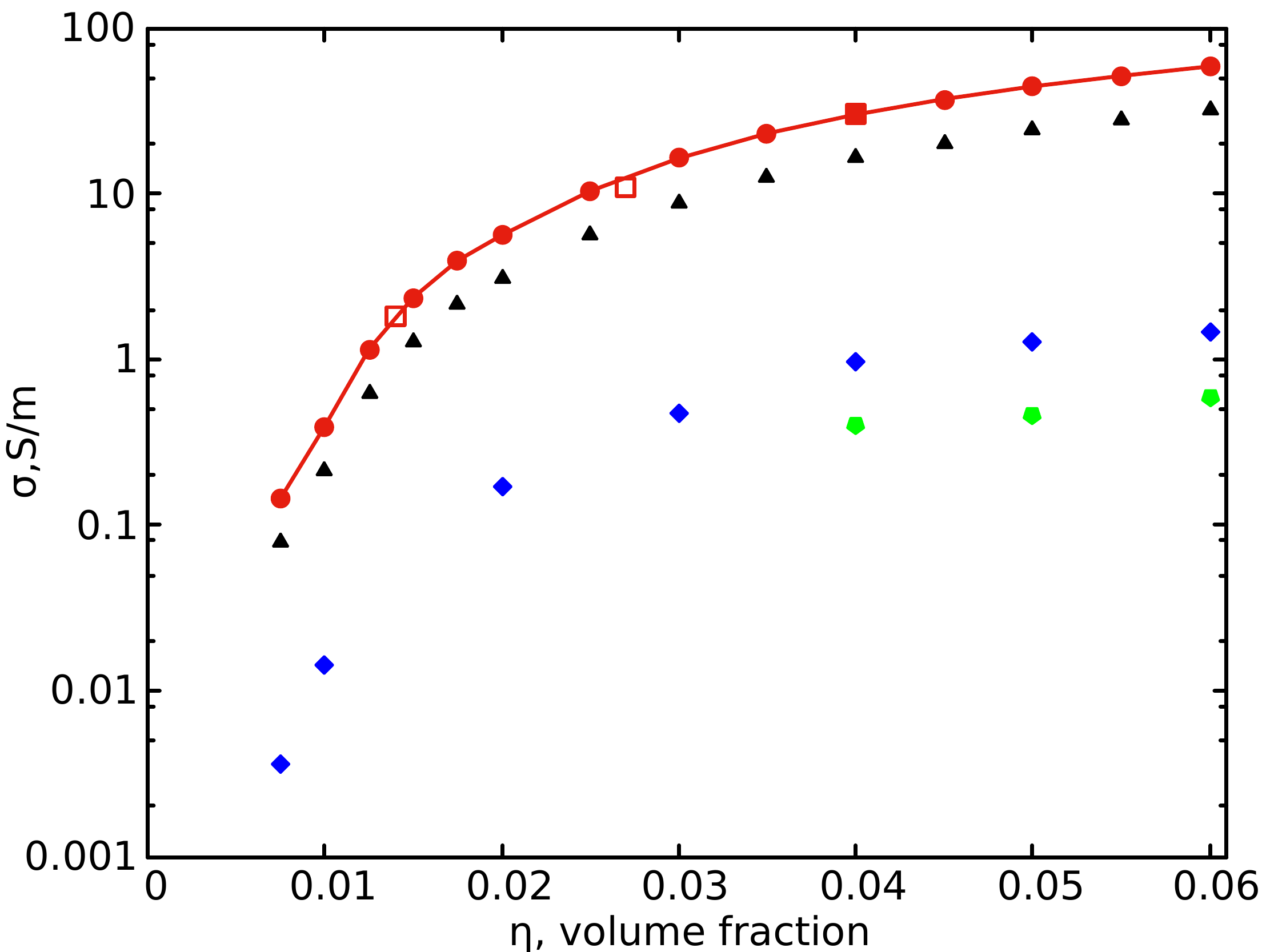}
\caption{ The conductivity of CNT enhanced nanocomposites above the percolation threshold  obtained in this work. The symbols of different shapes and colors are used to designate the following results. The red circles:  the fixed CNTs tunneling junctions resistance of $R=1$ M$\Omega$ is used, the red squares: the conductivity results for the fixed 1 M$\Omega$  tunneling resistance from \cite{Yu_2010}, the black triangles: the same as the red circles but for  $R=0.54$ M$\Omega$ corresponding to the mean value of the tunneling junction resistance for the parallel configuration from Table \ref{tab_cond}, the blue rhombi: the angle dependence of the CNTs junctions resistance is taken into account, the green pentagons: CNTs agglomeration is  considered in addition to the angle dependence. The red line is a guide for an eye.}
\label{fig_cond_pol}
\end{figure}

The dependence of the calculated composite conductivity $\sigma$ on CNTs volume fraction $\eta$ $\sigma(\eta)$ is presented in Fig. \ref{fig_cond_pol}. 
The value of the percolation threshold $\eta_{thresh}$ is estimated  in this work as $\eta_{thresh} = 0.007$. To test our realization of the percolation algorithm against the previous results of \cite{Yu_2010} we calculated the composite conductivity using
the fixed CNTs junction conductance equal to 1 M$\Omega$ for all junctions in a percolation network. Our results presented in
Fig. \ref{fig_cond_pol} by the red circles coincide within graphical accuracy to the results of \cite{Yu_2010} shown by the red squares.
The 1 M$\Omega$, used in various sources, for example, \cite{Yu_2010}, is not an arbitrary value, but rather a typical contact resistance of  CNTs junctions filled with polymer for simple geometries. In this work, we obtained  for the parallel configurations  $1/\langle G \rangle =0.54$ M$\Omega$. The $\sigma(\eta)$ dependence for the fixed tunneling resistance of $0.54$ M$\Omega$ is shown in Fig. \ref{fig_cond_pol} by the black triangles.

Taking into account the angle dependence of CNTs junctions conductances with the statistical parameters according to Table
\ref{tab_cond}, leads to the lowering of composite conductivity just above the percolation threshold by the factor of about 30. This number correlates  with the ratio of the mean conductances for the parallel, $\langle G \rangle_{par}$, and 45$^\circ$, $\langle G \rangle_{45}$, configurations: $f_G = \langle G \rangle_{par}/ \langle G \rangle_{45} = 53$, but is higher than $f_G$ due to the presence of junctions with $\varphi < \pi/4$.  

If agglomeration of CNTs, modeled by the inhomogeneity of their distribution according to formula (\ref{agglo}) and the parameter values discussed in section \ref{sec_percolation}, is taken into account in addition to the angle dependence of conductance, the composite conductivity is further reduced above a percolation threshold by the factor of 2.5. Lowering of conductivity of composites with agglomerated CNTs above a percolation threshold was also mentioned in \cite{Bao_2011}. The calculated results for the conductivity of a percolation network of agglomerated CNTs are shown in Fig. \ref{fig_cond_pol} by the green pentagons.

We believe that in this work we have identified some of the key factors that influence  nanocomposites electrical conductivity: the geometry of tunneling junctions and  changes of atomic configurations due to thermal fluctuations. Among other causes that may affect conductivity, but are not considered in this work,  are the presence of defects in CNTs, a distribution of CNTs over chiralities, lengths, aspect ratios, different separations between CNTs.

Until the specific experiments on conductivity for  R-BAPB polyimide are not available, we can make a preliminary comparison of our modeling results to the available experimental results for different composites. The calculated conductivity of composite just above the percolation threshold at $\eta=0.0075$ is equal to $3.6\cdot10^{-3}$ S/m.
This is a reasonable value that falls into the range of experimentally observed composites conductivities  (for the comprehensive compilation of experimental results see Table 1 of \cite{Eletskii}). To make a quantitative comparison of modeling results with experiments the full details of nanocomposites structure are necessary, including the CNTs parameters mentioned in the previous paragraph. All these factors can be easily incorporated into the  approach proposed in this work if sufficient computational resources are available.

\section*{Conclusions}

We have proposed a physically consistent, computationally simple, and at the same time precise, multi-scale method for calculations of electrical conductivity of CNT enhanced
nanocomposites. The method starts with the atomistic determination of the positions of polymer atoms intercalated between CNTs junctions, proceeds with the fully first-principles calculations of polymer-filled CNTs junctions conductance at the microscale and finally performs
modeling of percolation through an ensemble of CNTs junctions by the Monte-Carlo technique. 

The developed approach has been applied to the modeling of electrical conductivity of polyimide R-BAPB + single wall (5,5) CNTs nanocomposite.

Our major contributions to the field are the following.  We have proposed a straightforward
method to calculate a contact resistance and conductance for polymer-filled CNTs junctions with arbitrary atomic configurations without resorting to any simplifying assumptions. We have demonstrated that a consistent multiscale approach, based on solid microscopic physical methods can give reasonable results, lying within the experimental range, for the conductivity of composites and suggested a corresponding work-flow.

It is shown that a contact resistance
and nanocomposite conductivity is highly sensitive to the geometry of junctions, including an angle between CNTs axes and subtle thermal shifts of polymer atoms in an inter-CNT's gap.
Thus, we argue that for the precision  calculations of nanocomposites electrical properties rigorous atomistic quantum-mechanical approaches are indispensable.

We have to admit though, that we have not considered all possible factors that may influence CNT junctions conductivity on the micro-level. We concentrated on the CNTs crossing angle factor because it seems to be the
most influential. The additional factors may include, for example, defects in CNTs,
CNTs overlap lengths and others. On the other hand, the proposed approach may be used to take all those factors into account, provided sufficient computational resources are available.

\section*{Acknowledgments}

This work was supported by the State Program "Organization of
Scientific Research" (project 1001140) and by the Research Center "Kurchatov Institute" (order No.~1878~of~ 08/22/2019).

This work has been carried out using computing resources of the federal collective usage center Complex for Simulation and Data Processing for Mega-science Facilities at NRC “Kurchatov Institute”, \url{http://ckp.nrcki.ru/}.

\end{document}